# Improving Dynamic Range of Speckle Correlation based Optical Lever by spatial multiplexing

A. Vijayakumar[1,2,*], Shanti Bhattacharya[2], and Joseph Rosen[3]

[1] Centre for Micro-Photonics, Faculty of Science, Engineering and Technology, Swinburne University of Technology, Hawthorn VIC 3122, Australia.
[2] Department of Electrical Engineering, Indian Institute of Technology Madras, Chennai 600036, India.
[3] School of Electrical and Computer Engineering, Ben-Gurion University of the Negev, P.O. Box 653, Beer-Sheva 8410501, Israel.
*Corresponding author: vanand@swin.edu.au

## ABSTRACT

Speckle correlation based optical levers (SC-OptLev) possess attractive characteristics suitable for sensing small changes in the angular orientations of surfaces. In this study, we propose and demonstrate a spatial multiplexing technique for improving the dynamic range of SC-OptLev. When the surface is in its initial position, a synthetic speckle intensity pattern, larger than the area of the image sensor is created by transversely shifting the image sensor and recording different sections of a larger speckle pattern. Then, the acquired images are stitched together by a computer program into one relatively large synthetic speckle pattern. Following the calibration stage, the synthetic speckle intensity pattern is used to sense changes in the surface's angular orientation. The surface is monitored in real-time by recording part of the speckle pattern which lies within the sensor area. Next, the recorded speckle pattern is a cross-correlated with the synthetic speckle pattern in the computer. The resulting shift of the correlation peak indicates the angular orientations of the reflective surface under test. This spatial-multiplexing technique enables sensing changes in the angular orientation of the surface beyond the limit imposed by the physical size of the image sensor.

## Introduction

In 1826, Poggendorf invented the Optical Lever (OptLev) for improving the sensitivity of theodolites to an accuracy of 5 seconds of an arc and later, it was used by Gauss, Weber, Ising, Mol, and Burger for improving the accuracy of the measurements in their respective experiments.[1] Today, the OptLev is widely used for sensing angular variations of a mirror with a high accuracy.[2-9] In some cases, the OptLev is also used for the amplification of small displacements.[10] The enormous mirrors in the Laser Interferometer Gravitational-Wave Observatory (LIGO) are susceptible to angular tilts due to radiation pressure. OptLevs are therefore, also used in LIGO[11] and KAGRA[12] to detect any small changes in the angular orientation of the mirrors in order to correct them in real-time. In such sensitive tilt measurements,[12] homodyne detection is implemented, where the signal is extracted from a carrier wave by comparison with a reference. In a few studies, a quadrant photodiode was used as a sensor and from the differential current measurement, the shift was measured with a high accuracy.[13,14] One of the main drawbacks in using a quadrant photodiode (QPD) is that the dynamic range of sensing is limited by the small physical area of the sensor, which is usually only a few tens of micrometers. If the QPD is replaced by an image sensor such as a CMOS/CCD chip, then the dynamic range can be extended further. The optical configuration of a simplified OptLev is shown in Figure 1.

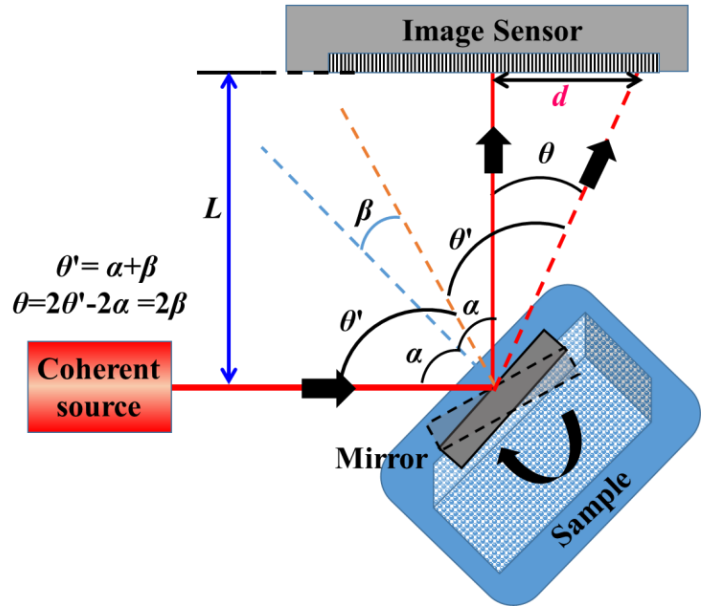

**Figure 1**: Optical configuration of a conventional OptLev. Light from a coherent source illuminates a mirror attached to observe the sample. The light reflected from the mirror is captured by the image sensor. Any variation in the angular orientation of the sample is measured from the shift of the optical beam on the image sensor.

Light from a frequency-stabilized coherent source is incident on a plane mirror, which is mounted with its surface at an angle of $\alpha$ with respect to the optical axis. As a result, the reflected light deviates at an angle of $2\alpha$ with respect to the incident light and is collected by an image sensor. The distance between the mirror and the image sensor is $L$. Since the mirror is attached to a sample whose angular orientation is under observation, any change in the angular orientation of the object would change the mirror tilt accordingly. When the sample is rotated by an angle $\beta$, the mirror is rotated by the same angle and the optical beam is deviated by an angle $\theta = 2\beta$. If the initial and final locations of the beam spot on the image sensor are known, then the spot displacement $d$ is related to the angular variation of the sample $\beta$ by the relation $d = L \cdot \tan(2\beta)$.

Assuming that the angle $\beta$ is small, the above equation reduces to $d = 2L\beta$, providing a linear relationship between $d$ and $\beta$. Therefore, the angular sensitivity of the measurement can be increased easily by increasing $L$. As a result, the conventional OptLev with a mirror and a sensor can be easily adapted to measure small variations with a high accuracy. For an image sensor with a pixel size of $\Delta$ and consisting of $m \times m$ *pixels*, the physical size of the image sensor is $\Delta \cdot m$ along the vertical and horizontal directions. The dynamic range of measurement of $\beta$ is $R = \pm 0.5\tan^{-1}(\Delta \cdot m/2L)$, which is the maximum displacement detectable by the image sensor of breadth $\Delta \cdot m$. The minimum angle which can be sensed by the OptLev when the detected signal is shifted by a full pixel is $\beta_{min} = (\Delta/2L)$. Since the sensitivity is defined as $S=1/\beta_{min}$, the dynamic range can be expressed as a function of the sensitivity as $R \approx \pm m/(2S)$. From this last equation, it is clear that when $R$ increases, $S$ decreases and vice versa, indicating the trade-off between the sensitivity and dynamic range.

In 1976, a tilt measurement technique using speckle correlation was introduced by Gregory.[15] This idea began to gain attention in the next few years.[16-21] A real-time surface inspection technique based on speckle correlation was proposed and demonstrated with an optical matched filter in 1992 by Hinsch *et.al.*[22] Several factors made speckle correlation more attractive compared to the interference techniques that had been used earlier for sensing tilt and deformation.[23-26] For example, speckle correlation required fewer optical elements, used an interferenceless optical configuration and gave rise to the possibility of real-time monitoring.[22] Alternatively, structured light techniques with spatial calibration were developed to measure tilts.[27] Currently, the speckle correlation technique is used as a sensing tool for various applications such as measurement of random processes in rough surfaces,[28,29] quantification of the corrosion process,[30] surface slope, deformation and motion measurements,[31-35] and sub-micrometer displacement measurement.[36]

Considering the disadvantages of interferometry such as the need for two beams and vibration isolation, the speckle correlation based OptLev (SC-OptLev) is chosen in this study. The SC-OptLev is compared to the conventional OptLev (C-OptLev)[13,14] with the help of Figure 2. Light from a coherent source illuminates a diffuser plate and the light scattered by the diffuser plate is incident on a plane mirror. The mirror is oriented with respect to the optical axis and deflects the scattered light such that the speckle pattern is recorded by the image sensor. When the mirror is tilted, the incident scattered light acquires a linear phase according to the change in the angular orientation of the mirror, propagates in a different direction and reaches the image sensor at a different lateral location. The scattered light from the diffuser is incident on the mirror and the complex amplitude after the mirror can be expressed as $A(x,y) \cdot \exp[j\Phi(x,y)]$, where $A(x,y) \in [0,1]$ and $\Phi(x,y) \in [0,2\pi]$. The light reflected from the mirror reaches the image sensor after propagating a distance of $L$ with a complex amplitude given by $A(x,y) \cdot \exp[j\Phi(x,y)] * Q(1/L)$, where $Q$ is the quadratic phase function given as $Q(a) = \exp[i\pi a\lambda^{-1}(x^2+y^2)]$ and '$*$' is a 2D convolution operator. The complex amplitude of light when the mirror undergoes an angular variation of $\beta$ along the $x$ direction is given as $A(x,y)\cdot\exp[j\{\Phi(x,y)+2\pi\lambda^{-1}\sin(2\beta)x\}]$ and the complex amplitude reaching the image sensor in this case is $A(x,y)\cdot\exp[j\{\Phi(x,y)+2\pi\lambda^{-1}\sin(2\beta)x\}]*Q(1/L)$. The intensity patterns recorded by the sensor before and after the angular variation of the mirror are $I_0 = |A(x,y)\cdot\exp[j\Phi(x,y)]*Q(1/L)|^2$ and $I_\beta = |A(x,y)\cdot\exp[j\{\Phi(x,y)+2\pi\lambda^{-1}\sin(2\beta)x\}]*Q(1/L)|^2$ respectively. The shift of the pattern on the image sensor is calculated by comparing the location of the correlation peaks obtained from $C_1 = I_0 \otimes I_0$ and $C_2 = I_0 \otimes I_\beta$, where '$\otimes$' is a 2D correlation operator. If the two intensity patterns $I_0$ and $I_\beta$ are zero-padded to double their size, the dynamic range of the C-OptLev is doubled.[37]

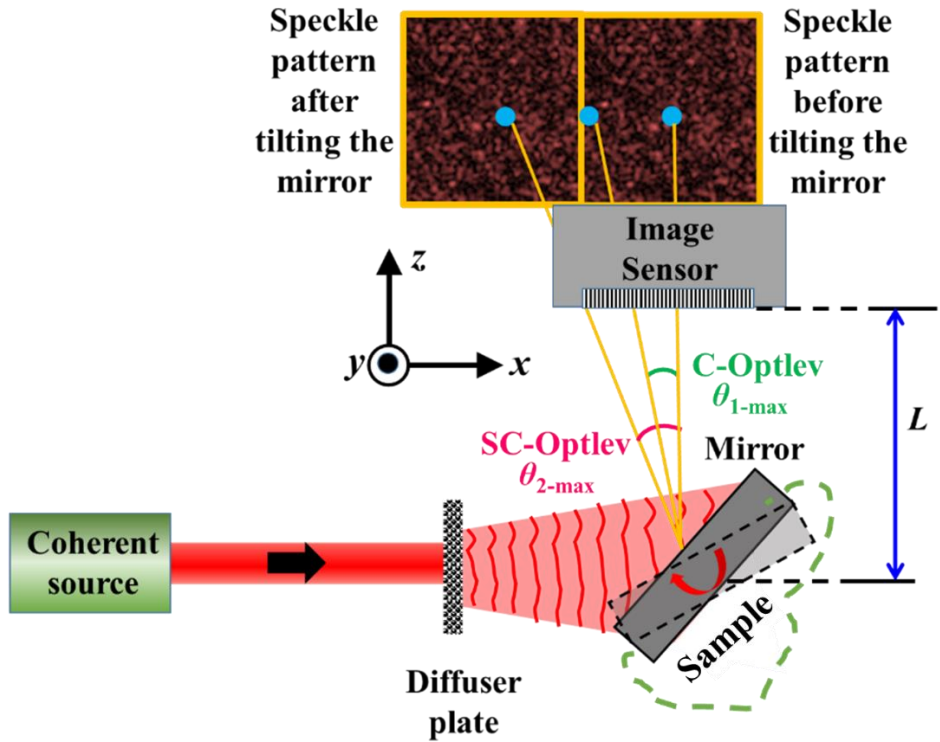

**Figure 2**: Optical configuration of an SC-OptLev. Light from a coherent source is scattered by a diffuser plate and the scattered light illuminates a mirror attached to the sample under observation. The light reflected from the mirror is captured by the image sensor. Any variation in the angular orientation of the sample is measured from the cross-correlation between the initial and final speckle intensity patterns.

SC-OptLev has various advantages compared to C-OptLev as the correlation process can extract not only tilt but also most of the surface deformations.[28-33] Even when the comparison is focused only on the measurement of angular variation of an sample, SC-OptLev exhibits a dynamic range twice as much as that of a C-OptLev.[37] Moreover, since the light from a coherent source such as a He-Ne laser has a beam diameter of few millimeters, it is often necessary to use a lens to focus the light to a smaller diameter on the sensor and therefore a C-OptLev is susceptible to lens aberrations. On the other hand, SC-OptLev requires simple components such as a diffuser plate, or an inexpensive thin scatterer. Under the assumption of space invariance, the cross-correlation used for the tilt measurement cancels out any aberrations induced by the optical components, since the same aberrations are expected in both recorded speckle intensity patterns. Furthermore, the width of the correlation peak can be engineered by signal processing techniques, and the measurement performances can be improved.[38] Therefore, it is advantageous to use an SC-OptLev compared to a C-OptLev for various applications. The accuracy of the two methods depends upon how narrow a focal spot or a correlation peak can be obtained for C-OptLev and SC-OptLev, respectively. Improvements in accuracy are out of the scope of the current study, and hence, are not discussed in this paper.

However, SC-OptLev like C-OptLev suffers from a trade-off between the dynamic range and sensitivity, although SC-OptLev has different limitations than C-OptLev in these two parameters. Moreover, in some optical configurations the width of the reconstructed correlation peak is dependent upon the scattering degree of the diffuser plate and the various distances of the optical setup.[39-41] There are other problems associated with an SC-OptLev such as the presence of background noise due to the correlation of speckle intensity patterns.[42,43] In this manuscript, we propose a spatial multiplexing technique to improve the dynamic range beyond the range of [37], without significantly affecting the sensitivity of the SC-OptLev. The presence of background noise due to the cross-correlation of speckle intensity patterns is also minimized herein. Note that it is not possible to monitor the angular variation in real-time since in SC-OptLev the observed signal is only a movement of a speckle pattern, unlike the movement of a focused spot in the case of C-OptLev. Various solutions are provided in [37] to address the real-time problem and to develop a SC-OptLev that is as simple as C-OptLev, but with a double the dynamic range.

## Methodology

The optical configuration of the spatial multiplexing SC-OptLev (SM-SC-OptLev) is shown in Figure 3. As mentioned above, light from a coherent source is incident on a diffuser plate, scattered and deflected by the surface whose angular tilt is to be monitored. The light deflected from the surface is recorded by an image sensor located at a distance of $L$ from the surface under test. The features of the speckle pattern on the sensor are dependent upon the scattering degree of the diffuser plate, its size and the distance $L$.[37] If $D$ is the size of the image sensor along the *x* and *y* directions, the maximum angular variation of the surface that can be detected using C-OptLev is $\beta_{1\text{-}max} = \pm 0.5\tan^{-1}(D/2L)$. However, under the assumption that the scattered light of the diffuser plate covers at least the entire image sensor, or more, in the initial state of no surface tilt, the maximum detectable angular variation in the case of SC-OptLev is $\beta_{2\text{-}max} = \pm 0.5\tan^{-1}(D/L)$.[37] Even when the speckle pattern is much larger than the sensor area, the maximum detection angle is still limited by the sensor area. In this section, a spatial multiplexing technique SM-SC-OptLev is proposed to overcome this limit.

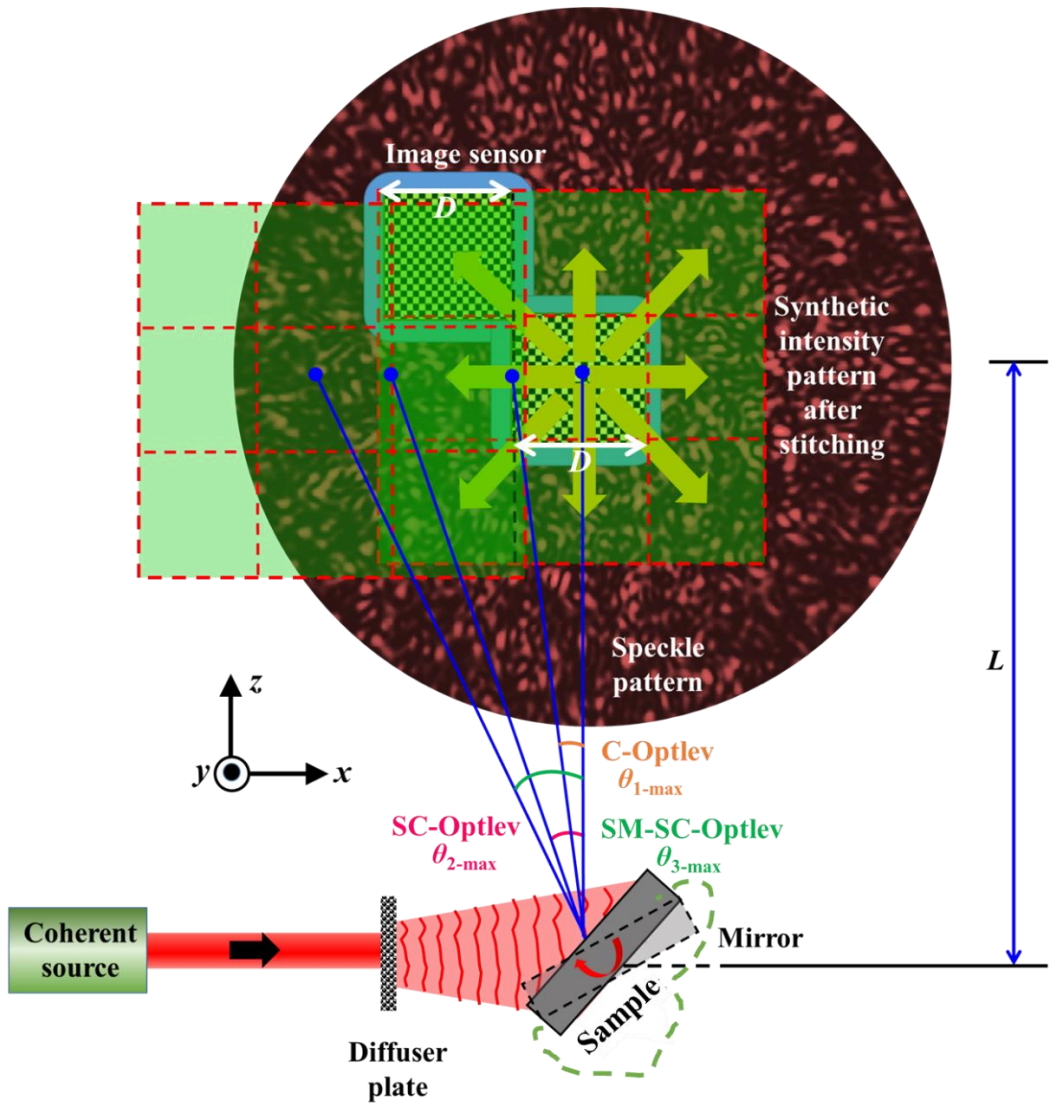

**Figure 3** Optical configuration of SM-SC-OptLev.

The proposed technique involves generation of a synthetic speckle intensity pattern, created by shifting the image sensor to different lateral locations and recording the speckle pattern at each location. Following the recording, the synthetic pattern is synthesized by a computational stitching procedure. If $(x_0, y_0)$ is the center of the image sensor and $I_0(x_0, y_0)$ is the speckle intensity pattern when the mirror is in its initial position, the part of the speckle intensity pattern recorded by the image sensor is given as $I_0(x_0,y_0)\text{Rect}[(x_0,y_0)/D]$. The synthetic speckle intensity pattern $I_{s0}$, with a size of $p$ times (*odd*) that of the image sensor, is expressed as

$$I_{s0} = \sum_{k=-\frac{p-1}{2}}^{\frac{p-1}{2}} \sum_{l=-\frac{p-1}{2}}^{\frac{p-1}{2}} I_0(x_0, y_0) Rect\left[\frac{(x_0 - Dk, y_0 - Dl)}{D}\right]$$

$$= I_0(x_0, y_0) Rect\left[\frac{(x_0, y_0)}{pD}\right]. \qquad (1)$$

When the mirror is tilted by an angle of $\beta$, the speckle intensity pattern shifts across the image sensor by a distance $d = L \cdot \tan(2\beta)$ and the shifted intensity pattern $I_\beta$ is recorded by the image sensor. $I_\beta$ is cross-correlated with the larger synthetic intensity pattern stitched computationally according to Eq. (1) from the $p$ recorded speckle patterns. The maximum tilt angle of the mirror along the $x$ and $y$ directions is $\pm 0.5 \tan^{-1}(pD/2L)$, where $p = 1, 3, 5\ldots$ The synthetic speckle intensity pattern stored in the computer is used as the reference pattern for the on-going tilt measurements. The speckle intensity pattern $I_\beta(x_0, y_0, t)\text{Rect}[(x_0, y_0)/D]$ recorded in real-time by the image sensor is zero-padded to match the size of the synthetic speckle pattern and is denoted by $I_\beta'(x_0, y_0, t)$. The location of the correlation spot is calculated by a 2D cross-correlation between $I_{s0}$ and $I_\beta'$. However, previous studies have shown that a correlation with the matched filter is not the ideal filter to obtain sharpest correlation peak due to the generated background noise. Alternatively, a phase-only filter is often implemented to obtain the sharp correlation peak $I_R$ with less background noise as given in Equation (2),[37,43-45]

$$I_R = \left|\mathcal{F}^{-1}\{exp[i\ arg(\tilde{I}_{s0})]|\tilde{I}_\beta'|exp[-i\ arg(\tilde{I}_\beta')]\}\right|, \qquad (2)$$

where $\tilde{I}_{s0} = \mathcal{F}(I_{s0})$ and $\tilde{I}_\beta' = \mathcal{F}(I_\beta')$, where $\mathcal{F}$ and $\mathcal{F}^{-1}$ are the Fourier transform and inverse Fourier transform operators, respectively. Recently, a non-linear adaptive correlation was implemented to reconstruct images with a background noise less than that of the phase-only filter.[46,47] The reconstructed image $I_R$ with a non-linear cross-correlation is given by

$$I_R = \left|\mathcal{F}^{-1}\{|\tilde{I}_{s0}|^o exp[i\ arg(\tilde{I}_{s0})]|\tilde{I}_\beta'|^b exp[-i\ arg(\tilde{I}_\beta')]\}\right|, \qquad (3)$$

where the values of $o$ and $b$ are tuned to minimize the background noise. In [46,47], entropy was used to find the values of $(o,b)$ that would reconstruct an object with the minimum background noise.

## Experiments

The spatial multiplexing technique is experimentally demonstrated using a set up similar to Figure 3. A *He-Ne* laser with a wavelength $\lambda = 632.8$ *nm* is used to illuminate a diffuser (HO-DF-25-X, Holmarc, *X*= 22 *μm*), where *X* is the grade of the diffuser. The light scattered by the diffuser is incident on a plane mirror that is oriented at an angle of 45º with respect to the optical axis and deflects the incident light by an angle of 90º with respect to the optical axis. The light deflected from the mirror was captured by an image sensor (Thorlabs Camera DCC1240M, 1024×768 pixels, pixel size: 4.65 *μm*). The distance between the mirror and the camera was 7 *cm,* while the distance between the diffuser and mirror was 3 *cm*.

The size of the image sensor is $4.76 \times 3.57$ *mm²*. For demonstration purposes only and to avoid mechanical movements, only the central part of the camera, roughly 1/25$^{th}$ of the total area *i.e.,* $0.95 \times 0.71$ *mm*, is used for recording the speckle intensity pattern. By doing so, the dynamic range is reduced to 1/5$^{th}$ of the initial value of the full image sensor. In the above optical configuration, for a C-OptLev, the dynamic range is decreased from $\beta_{x\text{-max}}= \pm 0.97º$ and $\beta_{y\text{-max}}= \pm 0.73º$ to $\beta_{x\text{-max}}= \pm 0.19º$ and $\beta_{y\text{-max}}= \pm 0.15º$. On the other hand, for an SC-OptLev, the dynamic range is decreased from $\beta_{x\text{-max}}= \pm 1.95º$ and $\beta_{y\text{-max}}= \pm 1.46º$ to $\beta_{x\text{-max}}= \pm 0.38º$ and $\beta_{y\text{-max}}= \pm 0.3º$. In this experiment, SM-SC-OptLev is implemented to sense variation in angular orientation of about five times higher than the limit imposed by the physical size of the image sensor. In the first step, a synthetic speckle intensity pattern is recorded with the mirror at an angle of 45º with respect to the optical axis by shifting the sensing area of 0.95×0.71 *mm* to different lateral locations on the camera plane, such that the entire $4.76 \times 3.57$ *mm* area of the speckle intensity pattern is recorded. One of the limitations of the SC-OptLev is that it is not possible to know when the angular limit has been reached, as only a motion of a speckle intensity pattern is seen on the camera. On the other hand, in C-OptLev, the spot disappears from the sensor clearly indicating that the limit has been reached. To overcome this problem associated with SC-OptLev, an open source based image acquisition procedure is proposed to convert the motion of the speckle pattern into the motion of a correlation peak in real-time.[37] Real-time monitoring of the correlation peak is also important in SM-SC-OptLev, in order to shift the location of the camera accurately while recording the synthetic speckle intensity pattern. A peak detection mechanism and real-time display of the angular orientation have been reported.[37]

The mirror was tilted manually to different orientations and the speckle intensity pattern with the size of $0.95 \times 0.71$ *mm* was recorded and zero-padded in the computer to the size of $4.76 \times 3.57$ *mm*. As verified from previous studies[37], a non-linear correlation is a more effective method compared to correlation with existing matched and phase-only filters. First, a non-linear correlation was executed between the synthetic speckle intensity pattern and the pattern recorded when the angular orientation of the mirror was varied along the *x* and *y* directions by $\beta_x = 0.25°$ and $\beta_y = 0.27°$. 121 different cross-correlations were computed for various values of *o* and *b* between -1 and 1 in steps of 0.2. The correlation results for the different values of *o* and *b* are shown in Figure 4. The optimal values with minimum entropy were found to be *o*=0.6 and *b*=0.2, where the entropy is defined as $S(o,b) = -\sum\sum \phi(m,n) log[\phi(m,n)]$, $\phi(m,n) = |C(m,n)|/\sum_M \sum_N |C(m,n)|$, *C(m,n)* is the correlation distribution, and *(m,n)* are the indexes of the correlation matrix.[46] The variation in the angular orientation of the mirror was determined from the distances where the correlation peak occurred between the autocorrelation peak ($x_0$=512, $y_0$=384) of $C_1$ and the cross-correlation peak ($x_\beta$=645, $y_\beta$=244) of $C_2$.

The speckle intensity patterns recorded for different angular orientations of the mirror and the cross-correlation results using non-linear correlation are shown in Fig. 5. It is seen from cases 5-7 that by SM-SC-OptLev, it is possible to detect the angular variation beyond the limit imposed by the physical dimension of the image sensor. The pixel location of the autocorrelation peak of the synthetic speckle intensity pattern is (512,384). The pixel locations of the cross-correlation peaks for the different angular variations of Figure 5 are given in Table 1. In the cases 1-4, the angular orientation of the mirror is within the limit of the image sensor while in the cases 5-7, the angular orientation of the mirror is greater than at least twice the size of the image sensor. In cases 2 and 3, the angular orientation of the mirror was varied only along the *x* and *y* directions respectively, while in case 4, the angular orientation of the mirror was varied along both *x* and *y* directions. In cases 5 and 6, the angular orientation of the mirror was again varied only along *x* and only *y* directions, respectively, but this time beyond the physical limits of the image sensor. Finally, in case 7, the angular orientation of the mirror was varied along both *x* and *y* directions beyond the physical limits of the image sensor. The correlation results of cases 5-7 for SM-SC-OptLev are indicative of the possibilities of sensing the angular orientation beyond the limits of C-OptLev as well as SC-OptLev.

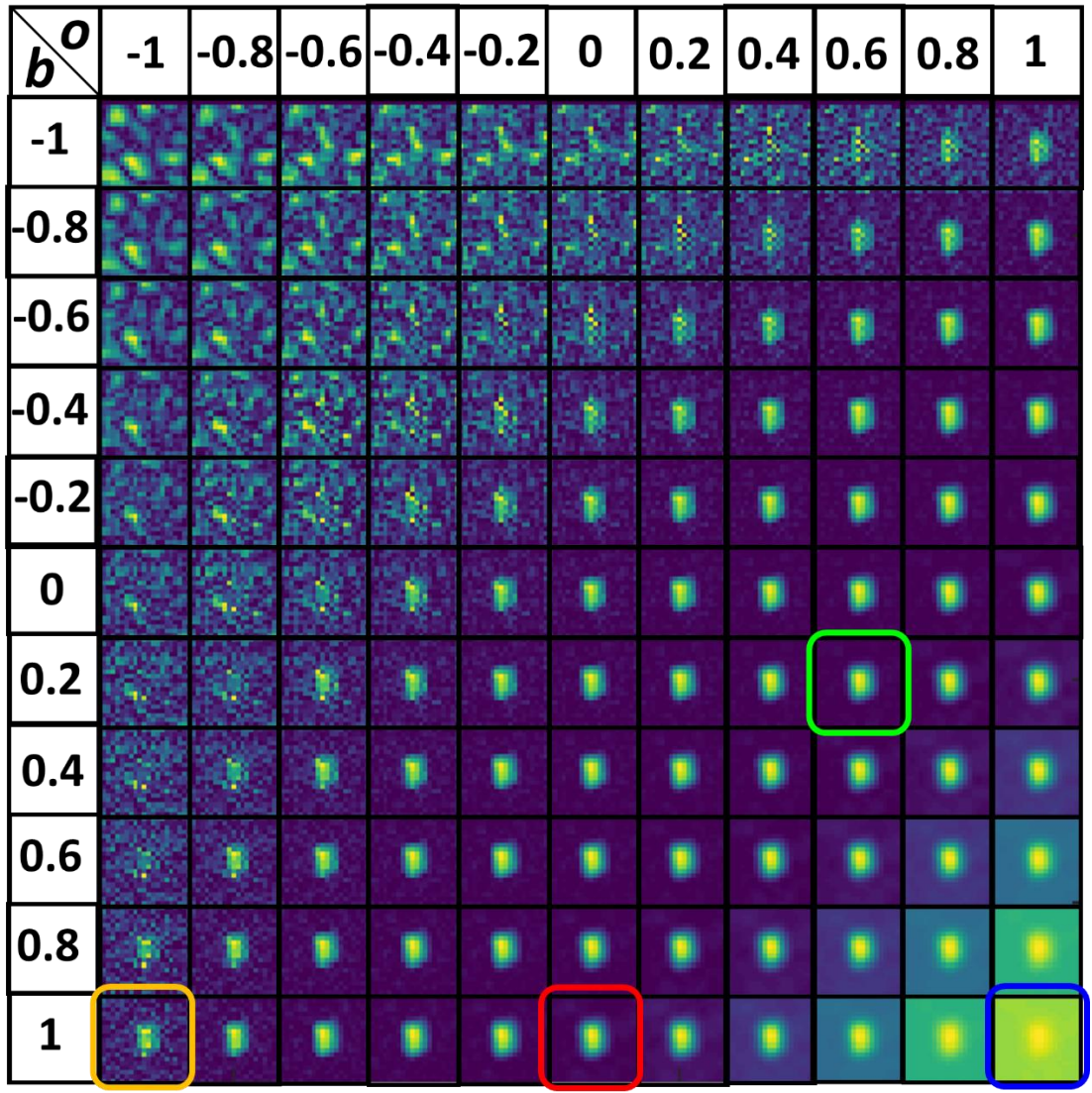

**Figure 4:** Correlation results of a non-linear correlator for different values of *o* and *b*. The results of inverse filter, matched filter, phase-only filter, and optimal filter are shown in yellow, blue, red and green boxes, respectively

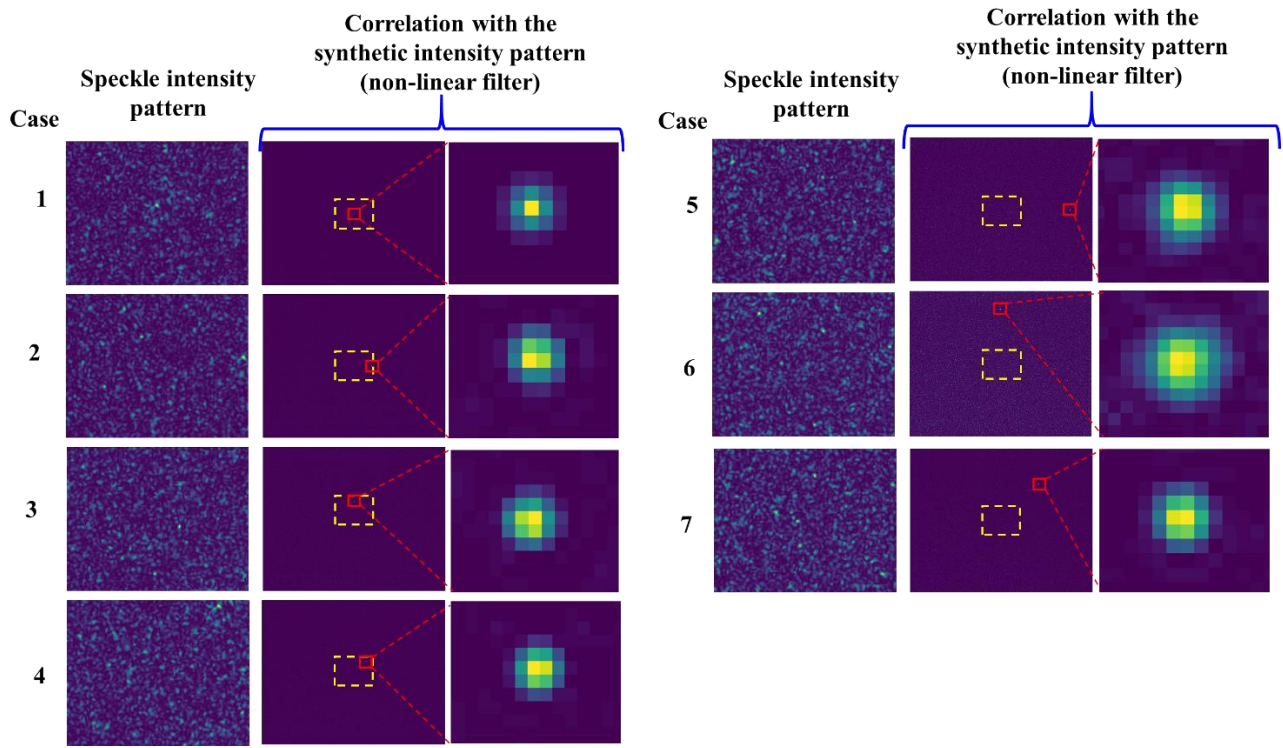

**Figure 5** Images of speckle intensity patterns for different tilt angles (before zero-padding) and cross-correlation functions obtained by non-linear correlation for different cases of the angular orientation of the mirror. The yellow boxes indicate the boundary of the image sensor.

From the experimental results of Fig. 5, it is seen that when the variation in the angular orientation is larger, there is an increase in the width of the correlation peak and decrease in the SNR. The experiment was repeated for different angular orientations of the mirror and the corresponding correlation peaks are plotted in Figure 6. The SNR is defined as 1/(Average background noise) and is plotted for the different angular orientation of the mirror. The SNR shown in Figure 7 is normalized using the highest SNR obtained for the zero variation of the angular orientation (autocorrelation). From Figure 6 and 7, it is indeed evident that with the increment of the tilt angle, the width of the correlation peak increases and the SNR decreases. However, the expected reduction in the sensitivity due to wider correlation peaks is by far lower than the sensitivity reduction obtained from the inverse relation of $S \propto R^{-1}$ mentioned in the introduction.

**Table – 1** Results of the 2D angular orientations. Initial pixel location of autocorrelation (512, 384), $L=7$ *cm* and camera pixel size = 4.65 *μm*.

| Case | Pixel Location (x) | Pixel Location (y) | Pixel Shift (x) | Pixel Shift (y) | Shift – x (*μm*) | Shift – y (*μm*) | $\beta_x$ (Deg) | $\beta_y$ (Deg) |
|---|---|---|---|---|---|---|---|---|
| 1 | 512 | 384 | 0 | 0 | 0 | 0 | 0 | 0 |
| 2 | 608 | 384 | 96 | 0 | 446.4 | 0 | 0.183 | 0 |
| 3 | 512 | 289 | 0 | 95 | 0 | 441.8 | 0 | 0.18 |
| 4 | 580 | 321 | 68 | 63 | 316.2 | 293 | 0.13 | 0.12 |
| 5 | 894 | 385 | 382 | 1 | 1776.3 | 4.65 | 0.727 | $2 \times 10^{-3}$ |
| 6 | 509 | 99 | 3 | 285 | 14 | 1325.3 | $6 \times 10^{-3}$ | 0.542 |
| 7 | 721 | 187 | 209 | 197 | 972 | 916 | 0.4 | 0.375 |

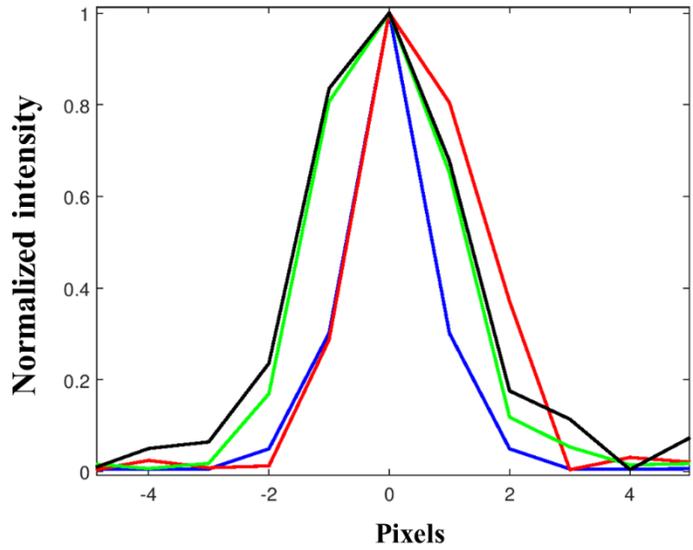

**Figure 6** Plots of the correlation peaks for a variation of the angular orientation along the *x* direction. The peaks are shifted by 0 µm (Blue), 446.4 µm (Red), 897.5 µm (Green) and 1348.5 µm (Black).

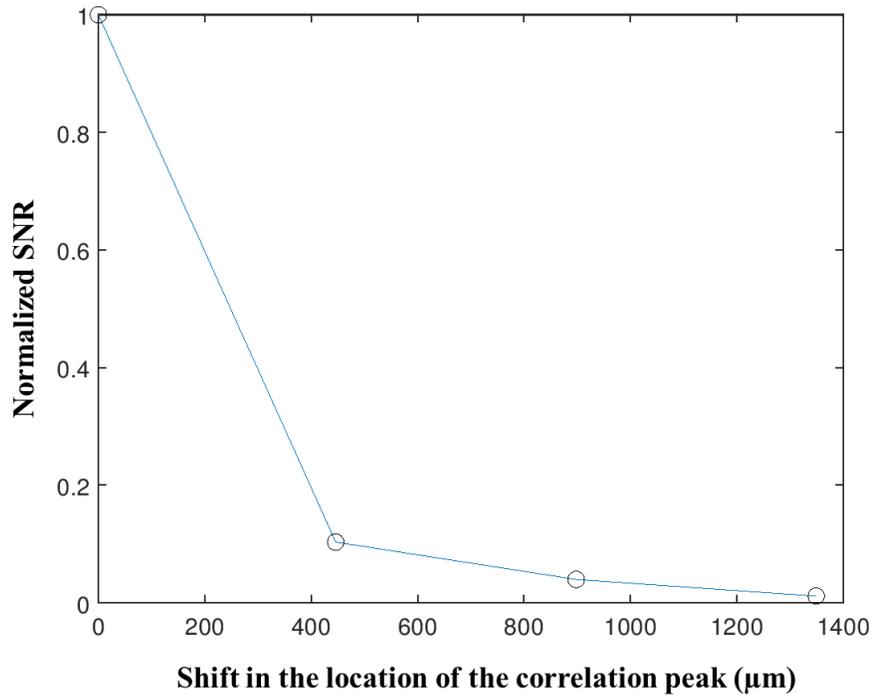

**Figure 7** Plot of the SNR with respect to the shift in the location of the correlation peak for 0 µm, 446.4 µm, 897.5 µm and 1348.5 µm.

## Conclusion

There are two important parameters namely dynamic range and sensitivity in monitoring the angular orientation of the sample under observation. We have proposed and demonstrated a spatial multiplexing technique for improving the dynamic range of the SC-OptLev – the dynamic range. In the spatial multiplexing technique, a synthetic speckle intensity pattern much larger than the size of the image sensor is generated and used as a reference pattern. It must be emphasized that this synthetic speckle intensity pattern is recorded only once and can be used to sense the variations of the angular orientation of the sample any number of times over any period of time. In this study, an optical configuration is constructed to exhibit a dynamic range enhancement of about 5 times that of a C-OptLev. In theory, there is no limit on how much the dynamic range can be increased. However, as much as the dynamic range is increased, it is expected that the sensitivity and the SNR will be reduced. Different filtering techniques have been studied in the past and it was concluded that non-linear correlation is the best candidate for reducing the noise.[37] In the SM-SC-OptLev, a non-linear filter is used to achieve maximum suppression of the background noise.

The advantage of SM-SC-OptLev is that the different areas of a larger speckle pattern are recorded only once followed by a computational stitching procedure. For real-time monitoring of the variations in the angular orientation, only one camera shot is required. We believe that the above advantage arises from the fact that a synthetic intensity pattern is used. The only disadvantage of SM-SC-OptLev is the one-time longer recording and processing time when the synthetic speckle intensity pattern is generated. Assistive optical and computational technologies using low-cost web camera and open source software have been developed for implementing an SC-OptLev for real-time monitoring of the variations in the angular orientation of the mirror.

In conclusion, the principle of spatial multiplexing has been implemented to improve the dynamic range without significantly reducing the sensitivity of an SC-OptLev. In other words, the existing trade-off between the dynamic range and sensitivity is relaxed such that the dynamic range has much larger ceiling and much smaller floor with a minor effect on the sensitivity. It must be stressed that the above technologies are not limited to sense only angles but can be used as powerful tools to sense deformations of various surfaces.

**Funding.** The authors thank LIGO R&D for India, IIT Madras and the Inter-University Centre for Astronomy and Astrophysics, India for funding this research. This study was done during a research stay of JR at the Alfried Krupp Wissenschaftskolleg Greifswald. This study was carried out when AV was a project officer in the Department of Electrical Engineering, IIT Madras.

## Author Contributions

## Additional Information

Competing financial interests: The authors declare that they have no competing interests.

All the figures were made by the authors.